\documentclass[preprint, 12pt, sort&compress]{elsarticle}

\usepackage{graphicx}%
\usepackage{multirow}%
\usepackage{amsmath,amssymb,amsfonts}%
\usepackage{amsthm}%
\usepackage{mathrsfs}%
\usepackage{mathtools}
\usepackage[title]{appendix}%
\usepackage{xcolor}%
\usepackage{textcomp}%
\usepackage{manyfoot}%
\usepackage{booktabs}%
\usepackage{algorithm}%
\usepackage{algorithmicx}%
\usepackage{algpseudocode}%
\usepackage{listings}%
\usepackage{hyperref}

\usepackage{comment}

\usepackage{color}
\usepackage{xcolor}

\begin{document}

	\begin{frontmatter}
		\title{Generative models struggle with kirigami metamaterials}
  
        \cortext[cor1]{Corresponding author}

        \author[1]{Gerrit Felsch}
        \author[1]{Viacheslav Slesarenko\corref{cor1}}    
        \ead{viacheslav.slesarenko@livmats.uni-freiburg.de}         
        \affiliation[1]{organization={Cluster of Excellence livMatS @ FIT – Freiburg Center for Interactive Materials and Bioinspired Technologies},
        addressline={Georges-Köhler-Allee 105},
        postcode={D-79110},
        city={Freiburg},
        country={Germany}}
		
        \begin{abstract}
        Generative machine learning models have shown notable success in identifying architectures for metamaterials — materials whose behavior is determined primarily by their internal organization — that match specific target properties. By examining kirigami metamaterials, in which dependencies between cuts yield complex design restrictions, we demonstrate that this perceived success in the employment of generative models for metamaterials might be akin to survivorship bias. We assess the performance of the four most popular generative models~—~the Variational Autoencoder (VAE), the Generative Adversarial Network (GAN), the Wasserstein GAN (WGAN), and the Denoising Diffusion Probabilistic Model (DDPM) — in generating kirigami structures. Prohibiting cut intersections can prevent the identification of an appropriate similarity measure for kirigami metamaterials, significantly impacting the effectiveness of VAE and WGAN, which rely on the Euclidean distance – a metric shown to be unsuitable for considered geometries. This imposes significant limitations on employing modern generative models for the creation of diverse metamaterials.
        
		\end{abstract}
		
	\end{frontmatter}

\section{Introduction}\label{sec1}

Mechanical metamaterials inherit unique behavior and extreme properties from their intricate internal organization. Conceptualized back at the end of the previous century \cite{lakes_foam_1987, milton_which_1995}, mechanical metamaterials underwent significant progress over the last thirty years \cite{qi_recent_2022, slesarenko_bumpy_2024}, in part thanks to the revolution in additive manufacturing, enabling precise fabrication \cite{askari_additive_2020, surjadi_mechanical_2019}. If classical metamaterials were mostly lattices assembled from the repeating unit cells \cite{zheng_ultralight_2014}, modern metamaterials often feature heterogeneous designs \cite{kumar_inverse-designed_2020}, exhibit multistability \cite{shan_multistable_2015} and are even capable of performing simple computations \cite{waheed_boolean_2020, jiao_mechanical_2023}. Such advanced mechanical behavior relies on structure-properties relationships more complex than ever before. Tailoring metamaterial architectures to achieve specific desired behavior is a challenging task, but it has become more accessible thanks to the advancements in artificial intelligence (AI) and machine learning (ML). A variety of machine learning approaches have gained recognition for their ability to rapidly create a large variety of different metamaterial designs once they have been trained \cite{zheng_deep_2023}. This includes Variational Autoencoders (VAE) \cite{Kingma2013AutoEncoding}, Generative Adversarial Networks (GAN) \cite{Goodfellow2014GenerativeAdversarial} and fairly recent the powerful Denoising Diffusion Models \cite{Bastek2023InverseDesign} that have revolutionized image generation \cite{Dickstein2015DeepUnsupervised, ho2020denoising}. 

However, upon examining the application of generative models in the inverse design of mechanical metamaterials \cite{zheng_deep_2023, lee_data-driven_nodate}, it is apparent that they are used differently than in image generation. In image generation, the strength of these models is in their ability to narrow the design space from all possible images to a subset of admissible ones, such as generating only human faces \cite{Liu2015faceattributes}. Contrary to this, it is common to parameterize mechanical metamaterials such that every randomly chosen set of parameters is admissible \cite{Bastek2022InvertingThe, Felsch2023ControllingAuxeticity}, or the restrictions are trivial, such as requiring the trusses to have positive lengths. For imposing more complex restrictions on the design space, advanced generative ML techniques are required. For example, graph-based deep learning generative framework enabling the construction of continuous latent space representation for an extreme variety of trusses has been recently presented \cite{zheng_unifying_2023}. Nonetheless, ML methods applied to metamaterials are primarily used to learn the relationship between structure and properties rather than to understand structural restrictions that should be applied. 

\begin{figure}[t]
        \centering
    	\includegraphics[width=\textwidth]{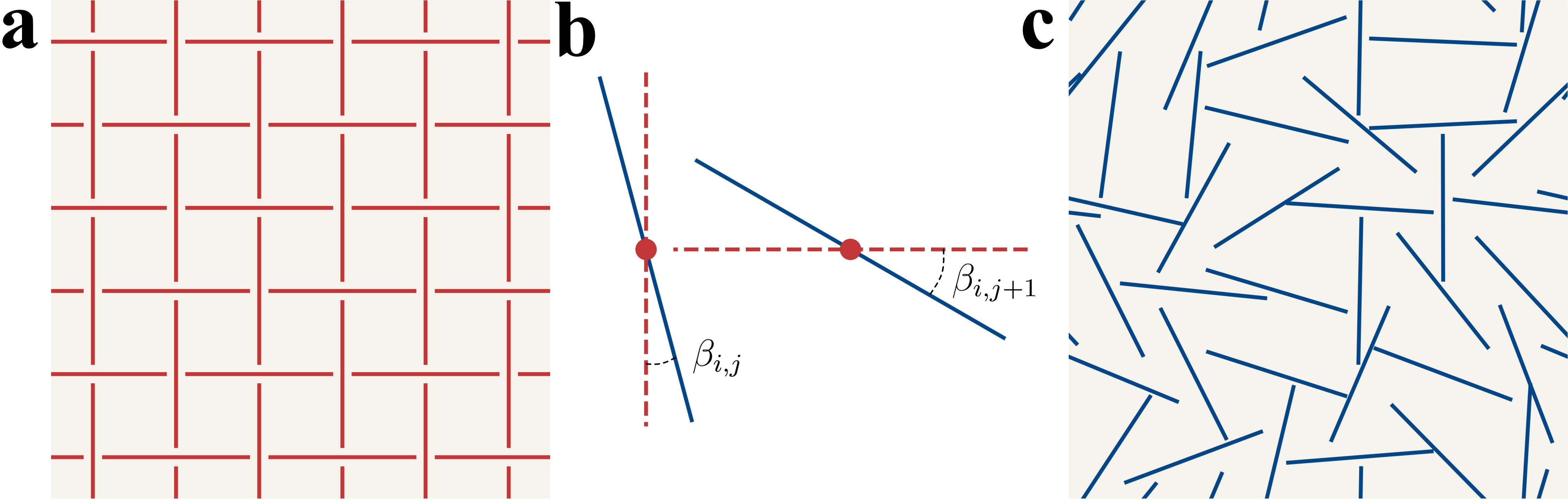}

		\caption{\text{Mechanical metamaterials based on straight cuts.} \textbf{a} Initial $6\times6$ pattern with alternating cuts. This architecture gives rise to auxetic behavior through the rotating squares mechanism. \textbf{b} Perturbation of the initial architecture is performed by adding rotations $\beta_{i,j}$ to each cut. The absolute value of rotations is capped by parameter $\beta_{max}$. \textbf{c} Resulting admissible design without intersections between cuts. The likelihood of obtaining intersection-free sample through random rotations ($\beta_{max}=90^\circ$) does not exceed 0.001\%}
		\label{fig:rotation}
\end{figure}

However, such all-admissible parameterization can come at the cost of excluding more sophisticated and sometimes optimal solutions from the design space. A good example is the metamaterials employing straight cuts in the planar sheets -- often called kirigami metamaterials \cite{jin_engineering_nodate, zhai_mechanical_2021} -- to program the desired mechanical behavior. For instance, alternating cuts with horizontal and vertical orientations within the sheet (Figure~\ref{fig:rotation}a) can lead to the manifestation of auxetic behavior via the rotating squares mechanism \cite{grima_auxetic_2000}. Further, it was shown \cite{Grima2016AuxeticPerforated} that adding random rotations to the base alternating structure could be a method to program the behavior of the resulting metamaterial. The greater the maximum deviation from the base structure, the greater the achievable range of material properties. It is important to note that rotations for each cut were limited to prevent intersections that can cause undesired effects such as stress concentration or even lead to disconnected regions. While limiting the deviations from the initial structure helps in preventing intersections, it simultaneously disqualifies the majority of intersection-free configurations, such as the one shown in Figure~\ref{fig:rotation}c.   

While kirigami metamaterials have been the focus of multiple studies employing machine learning techniques, such studies were either limited to property predictions \cite{liu_predictive_2020}, or employed strict restrictions to get all-admissible parameterizations \cite{hanakata_forward_2020, alderete_machine_2022}. This prompts the question of why ML is not more extensively used to learn design restrictions in metamaterials in a manner akin to its application in image generation. In this manuscript, we examine the fundamental differences between image and metamaterial generation. Additionally, we discuss why certain ML algorithms can learn these design restrictions while others cannot. Through this analysis, we aim to highlight the presence of survivorship bias in the existing literature on generative AI in metamaterial design, which arises from considering only metamaterials with favorable design spaces. Furthermore, we demonstrate that the challenge of generating the aforementioned kirigami metamaterials serves as an effective benchmark for assessing the ability of generative algorithms to learn general design space restrictions.

\section{Problem statement}

Most generative design algorithms adhere to a similar core concept. A dataset filled with examples of what to generate, e.g. admissible metamaterials, is presented to the corresponding algorithm that learns from this information to generate similar data. The difficulty lies in determining what qualifies as "similar data". There are two approaches to this, which can also be combined together. The first approach views the dataset as samples from a probability distribution, where certain combinations of parameters are more or less likely than others. In this case, similarity is assessed by comparing the distribution of the generated data with that of the example data. The second approach performs a direct comparison between samples. For this, a specific sample-to-sample similarity metric must be chosen. The most commonly used one is the Euclidean distance (ED). In a n-dimensional space $\mathbb{R}^n$, the ED between two samples $x$ and $\hat x$ is calculated as follows:

\begin{equation}
    D_E(x, \hat x) =   \sqrt{\sum_{i=1}^n (x_i-\hat x_i)^2} 
\end{equation}

\begin{figure}[t]
    	\centering
     \includegraphics[width=0.9\textwidth]{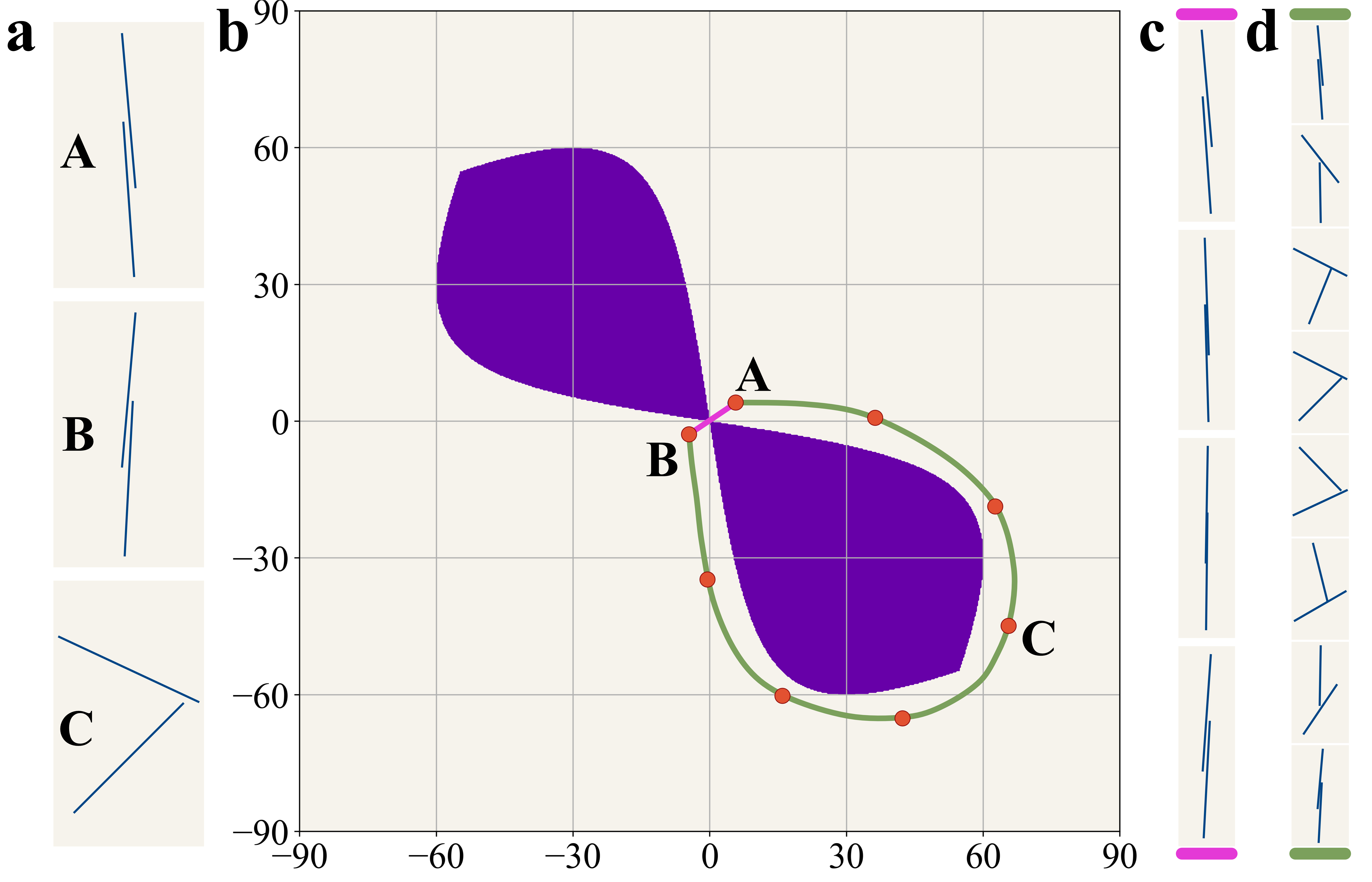}
		\caption{\text{Suitability of the Euclidean Distance for two cuts}. \textbf{a} Three different configurations (\textbf{A:} [$5^\circ$,$4^\circ$], \textbf{B:} [$-5^\circ$,$-3^\circ$] \textbf{C:} [$65^\circ$,$-45^\circ$]) of adjacent cuts with unit length between centers and length of $\sqrt{3}$. \textbf{b} The design space for the considered system with two cuts. Dark blue zones correspond to the angle pairs of intersecting cuts. Magenta and green lines show two possible routes between \textbf{A} and \textbf{B}. \textbf{c} Sequence of cut positions corresponding to direct transition from \textbf{A} and \textbf{B} (magenta path). \textbf{d} Sequence of cut positions for detour path shown by green line. Note passing configuration \textbf{C} on a route from \textbf{A} to \textbf{B}.}
		\label{fig:configurations}
\end{figure}	

 While other similarity measures might be more effective for image generation \cite{Adler2018Advances}, ED has still been shown to yield good results, and it is a part of the original formulations for several generative algorithms  \cite{Kingma2013AutoEncoding, Arjovsky2017WassersteinGenerative}. For mechanical metamaterials, on the other hand, ED may not be the most appropriate choice for measuring similarity. For illustration, consider Figure~\ref{fig:configurations}a, which displays three different configurations for two neighboring cuts with angles to the vertical direction as follows: \textbf{A)} [$5^\circ$,$4^\circ$], \textbf{B)} [$-5^\circ$,$-3^\circ$] \textbf{C)} [$65^\circ$,$-45^\circ$]. We can pose the question: which two configurations are most similar? At first glance, the answer (\textbf{A}, \textbf{B}) appears straightforward, which aligns with ED since $D_E(\textbf{A},\textbf{B})<D_E(\textbf{A},\textbf{C})<D_E(\textbf{B},\textbf{C})$, where

\begin{equation}
    \begin{split}
        D_E(A,B) &= \sqrt{(5+5)^2 + (4+3)^2} \approx 12.206\\
        D_E(A,C) &= \sqrt{(5-65)^2 + (4+45)^2} \approx 77.466 \\
        D_E(B,C) &= \sqrt{(-5-65)^2 + (-3+45)^2} \approx 81.633\\
    \end{split}
\end{equation}

However, this conclusion overlooks a crucial point. Dark regions in Figure~\ref{fig:configurations}b represent pairs of angles where two neighboring cuts intersect, forming a non-admissible zone. The transition from configuration \textbf{A} to configuration \textbf{B} through linear interpolation (as illustrated in Figure~\ref{fig:configurations}c) follows the shortest path in Euclidean space, indicated by the magenta line. This path, however, clearly passes through the non-admissible zone.  Consequently, not all intermediate configurations between \textbf{A} and \textbf{B} belong to the admissible design space, despite both end configurations being intersection-free. To navigate from \textbf{A} to \textbf{B} staying within the admissible design space, a considerably longer trajectory is required, as shown by the green line in Figure~\ref{fig:configurations}b. Notably, the admissible path from \textbf{A} to \textbf{B} includes passing through configuration \textbf{C} (Figure~\ref{fig:configurations}d). Therefore, if no intersections are allowed, the title of most similar pair belongs to (\textbf{A}, \textbf{C}). This implies that ED might not be an appropriate measure of similarity for these simplified two-cut kirigami designs. Moreover, it suggests that generative algorithms relying on ED may not effectively learn to avoid intersections, which will be shown further. This paper analyzes the ability of the four most-common generative design algorithms -- VAE \cite{Kingma2013AutoEncoding}, GAN \cite{Goodfellow2014GenerativeAdversarial}, Wasserstein GANs (WGAN) \cite{Arjovsky2017WassersteinGenerative} and Denoising Diffusion Probabilistic Models \cite{ho2020denoising} -- to handle such geometrical challenges.

\begin{figure}[t]
        \centering
    	\includegraphics[width=\textwidth]{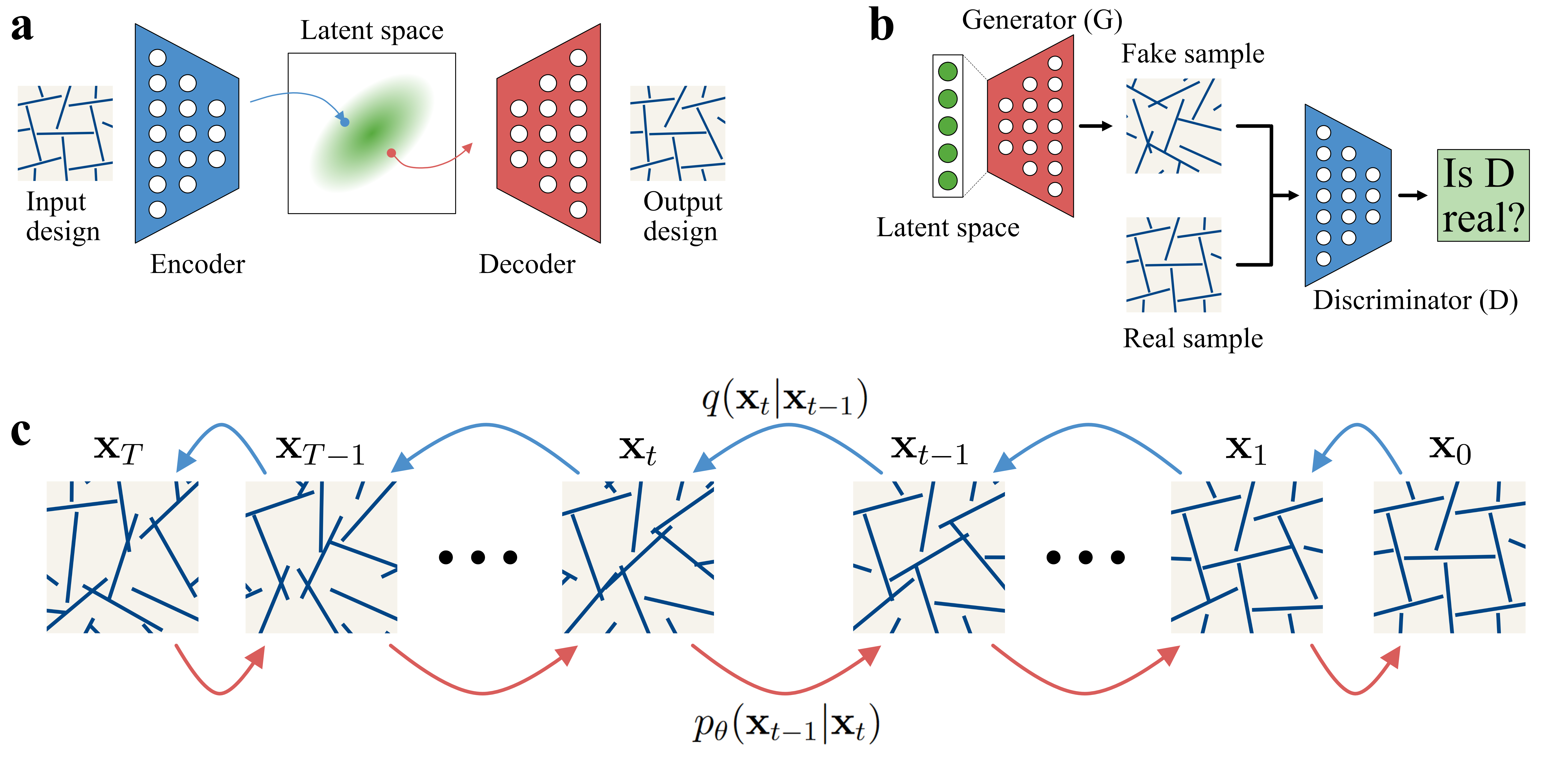}

		\caption{\text{Generative approaches}. \textbf{a} Variational Autoencoder (VAE), comprised of Encoder and Decoder stages, learns to map the designs into latent space and retrieve them back. \textbf{b} Generative Adversarial Network (GAN) utilizes competition between Generator and Discriminator to create samples that look real. \textbf{c} Denoising Diffusion Probabilistic Model (DDPM) employs sequential addition of noise to map the designs to latent space.}
		\label{fig:algorithms}
\end{figure}	

\subsection{Variational Autoencoders}
The Variational Autoencoder (VAE), shown in Figure~\ref{fig:algorithms}a and introduced in 2013 by Kingma and Welling \cite{Kingma2013AutoEncoding}, is a modification of the traditional autoencoder \cite{Ballard1987ModularLearning} for generative design. Autoencoders transform input data into a usually lower-dimensional representation in so-called latent space and are comprised of two parts. The encoder creates a representation of the original data in the latent space while the decoder tries to reconstruct the original data from such a representation. Both parts are trained jointly by minimizing the error between the original and reconstructed data, which is usually referred to as reconstruction loss. While these traditional autoencoders can be used for a variety of tasks, including image denoising \cite{Vincent2008ExtractingAndComposing}, dimensionality reduction \cite{Hinton2006ReducingTheDimensionality}, and anomaly detection \cite{Sakurada2014AnomalyDetection} they lack the ability to generate new data. Unlike traditional autoencoders, VAEs map to mean and variance of a normal distribution in the latent space. An additional regularization term in the loss function compels this distribution to have zero mean and unit variance. This enables the generation of new data by sampling a representation from this normal distribution and then decoding it. While the Kullbeck-Leiber divergence \cite{Kullback1951OnInformation} has been consistently used as regularization loss, the choice of reconstruction loss, in general, depends on the input data. In the original formulation, Binary Crossentropy loss was used for the MNIST dataset, where pixels are supposed to be either white or black, while the Mean Squared Error (the square of the ED) was used for continuous problems \cite{Kingma2013AutoEncoding}. Together they yield the following combined loss function:

\begin{equation}
    \mathcal{L}_{VAE} = \underbrace{D_{KL} \big(\mathcal{N}(\mu_x, \sigma_x), \mathcal{N}(0, 1) \big)}_\text{regularization loss} + \underbrace{\kappa D_E(x,\hat{x})^2}_\text{reconstruction loss}
\end{equation}
, where $x$ is the original input, $\hat{x}$ the reconstructed one and $\mu_x, \sigma_x$ are the mean and variance of the learned distribution in the latent space, while $\kappa$ is a parameter controlling the trade-off between the two losses. Note the reliance of reconstruction loss on the Euclidean distance. We note that there have been efforts to make VAEs independent of the ED, such as by combining it with a GAN, that have shown promising results \cite{Larsen2016Autoencoding}.

\subsection{Generative Adversarial Networks}
In 2014, Goodfellow et al. \cite{Goodfellow2014GenerativeAdversarial} introduced a framework for training generative models via an adversarial process, giving rise to Generative Adversarial Networks (GANs). This architecture (Figure~\ref{fig:algorithms}b) encompasses both a generative model $\mathcal{G}$, and a discriminative model $\mathcal{D}$, which are trained simultaneously. This training takes the form of a two-player game. While $\mathcal{D}$ is trained to distinguish between data generated by $\mathcal{G}$ and the training data, $\mathcal{G}$ is simply trained to maximize the probability of $\mathcal{D}$ making a mistake. This process reaches an equilibrium when the generative model has learned a mapping between a chosen distribution in the latent space and the data distribution, similar to the discriminator of a VAE. It has further been shown that it is equivalent to minimizing the Jensen-Shannon divergence between the distribution of the data generated by $\mathcal{G}$ and the distribution of the training data:

\begin{equation}
    \mathcal{L}_{GAN} = D_{JS} \big(p(x), p(\hat x) \big) = D_{KL} \big(p(x), p(\hat x) \big) + D_{KL} \big(p(\hat x), p(x)\big) 
\end{equation}

This means that the vanilla GAN relies only on distances between probability distributions and not on the distance between samples, which should allow it to learn intersection avoidance.

\subsection{Wasserstein Generative Adversarial Networks}
The Wasserstein Generative Adversarial Network (WGAN) \cite{Arjovsky2017WassersteinGenerative} is a GAN variant where the Jensen-Shannon divergence has been replaced by the Wasserstein distance (also called Kantorovich–Rubinstein metric or Earth mover distance) to measure the discrepancy between the distributions of the generated and the training data. This metric, which was first introduced by Kantorovich \cite{Kantorovich1960MathematicalMethods}, is based on the principle of optimal transport. A probability distribution is seen as a distribution of mass in the design space, and the difference between distributions is measured as the minimal cost of transporting mass so that one distribution resembles the other. While it has the benefit of always staying finite and can, therefore, always provide meaningful gradients to update the generator, the Wasserstein distance ($\text{Wass}_1$) relies on an underlying metric to measure how far the mass has been transported. For the WGAN, this is usually the Euclidean metric $D_E$:

\begin{equation}
    \text{Wass}_1(p(G(z)),p(x)) = \inf_{\pi \in \Pi(G(z)),p(x))} \mathbb{E}_{(X_1,X_2) \sim \pi} D_E(X_1,X_2)
\end{equation}

Note that replacing the Jensen-Shannon by the Wasserstein distance makes training generally more stable, but WGAN formulation comes with the drawback of not always converging to the equilibrium point \cite{Mescheder2018WhichTraining}.

\subsection{Denoising Diffusion Probabilistic Models}

In 2015, Sohl-Dickstein \textit{et al.} \cite{Dickstein2015DeepUnsupervised} laid the foundation for a class of latent variable models, which have since become widely known as Denoising Diffusion models (Figure~\ref{fig:algorithms}c). The ingenious concept behind these models is that, rather than attempting to learn an arbitrary direct mapping from the latent space to the design space, they learn to reverse a diffusion process that stepwise transforms an image into its representation in the latent space. This diffusion process is defined as a Markov Chain - a stochastic model where the probability of transitioning to another state depends only on the current state - that gradually introduces noise to the image over a number of steps $T$. So for data of the from $\textbf{x}_0 \sim q(\textbf{x}_0)$ the forward process $q(\textbf{x}_{1:T}|\textbf{x}_0)$ is given as:
\begin{equation}
    q(\textbf{x}_{1:T}|\textbf{x}_0) = \prod_{t=1}^{T} q(\textbf{x}_t|\textbf{x}_{t-1})
\end{equation}
where the variance of the Gaussian noise that is added in each step is altered for each step based on a variance schedule $\beta_1, \ldots,\beta_T$:
\begin{equation}
    q(\textbf{x}_t|\textbf{x}_{t-1}) \coloneqq \mathcal{N}(\textbf{x}_t;\sqrt{1-\beta_t}\textbf{x}_{t-1},\beta_t\textbf{I})
\end{equation}
This formulation possesses the advantage that when the variances of the forward process $\beta_t$ are small, the reverse process $ p_{\theta}(\textbf{x}_{0:T}) $ can also be described as a Markov chain with Gaussian transitions. Only that in this instance, both the mean and variance of the transitions are learned \cite{Dickstein2015DeepUnsupervised}:
\begin{equation}
    p_\theta(\textbf{x}_{1:T}) = \prod_{t=1}^{T} p_\theta(\textbf{x}_{t-1}|\textbf{x}_{t})
\end{equation}
where
\begin{equation}
    p_\theta(\textbf{x}_{t-1}|\textbf{x}_{t}) \coloneqq \mathcal{N}(\textbf{x}_{t-1};\boldsymbol{\mu}_\theta(\textbf{x}_t,t), \boldsymbol{\Sigma}_\theta(\textbf{x}_t,t)), \quad p(\textbf{x}_T) = \mathcal{N}(\textbf{x}_T;\boldsymbol{0}, \textbf{I})
\end{equation}

Training of the reverse process is usually performed by minimizing the variational upper bound on the negative log-likelihood using stochastic gradient descent. In 2020 Ho \textit{et al.} \cite{ho2020denoising} introduced a variant of these models called Denoising Diffusion Probabilistic Model (DDPM). When conditioned on $\textbf{x}_0$, this bound can be rewritten using KL divergences between Gaussians:
\begin{equation}
    \begin{split}
    L(\theta) = \mathbb{E}_q \Big [ & D_{KL}( q(\textbf{x}_T | \textbf{x}_0) \|  p(\textbf{x}_T)) + \sum_{t>1} D_{KL}( q(\textbf{x}_{t-1} | \textbf{x}_t, \textbf{x}_0) \|  p_\theta(\textbf{x}_{t-1} | \textbf{x}_t)) \\ &- \log p_\theta(\textbf{x}_0 | \textbf{x}_1) \Big]
    \end{split}
\end{equation}

Furthermore, as these KL-divergences are computed between two Gaussian distributions, reparametrization enables the formulation of the variational bound as the MSE between the actual noise $\epsilon \sim \mathcal{N}(\textbf{0},\textbf{I})$ and its predicted counterpart $\epsilon_\theta(\textbf{x}_t,t)$:
\begin{equation}
    L(\theta) \coloneqq \mathbb{E}_{t,\textbf{x}_0,\epsilon} \Big[ \| \epsilon - \epsilon_\theta(\textbf{x}_t,t) \|^2 \Big]
\end{equation}
Here, once again, ED is encountered, though this time it is computed between instances of noise rather than instances from the design space. As this distance is derived from the initial choice of Gaussian noise in the forward process, it means that this choice limits the ability of denoising diffusion models to deal with non-Euclidean data. Therefore, research to extend diffusion models has recently focused on more general choices of forward processes and their reversal \cite{DeBortoli2022Advances, Huang2022Riemannian, Zhuang2023diffusion}.

\section{Kirigami dataset}

\begin{figure}[t]
        \centering
        \includegraphics[width=0.9\textwidth]{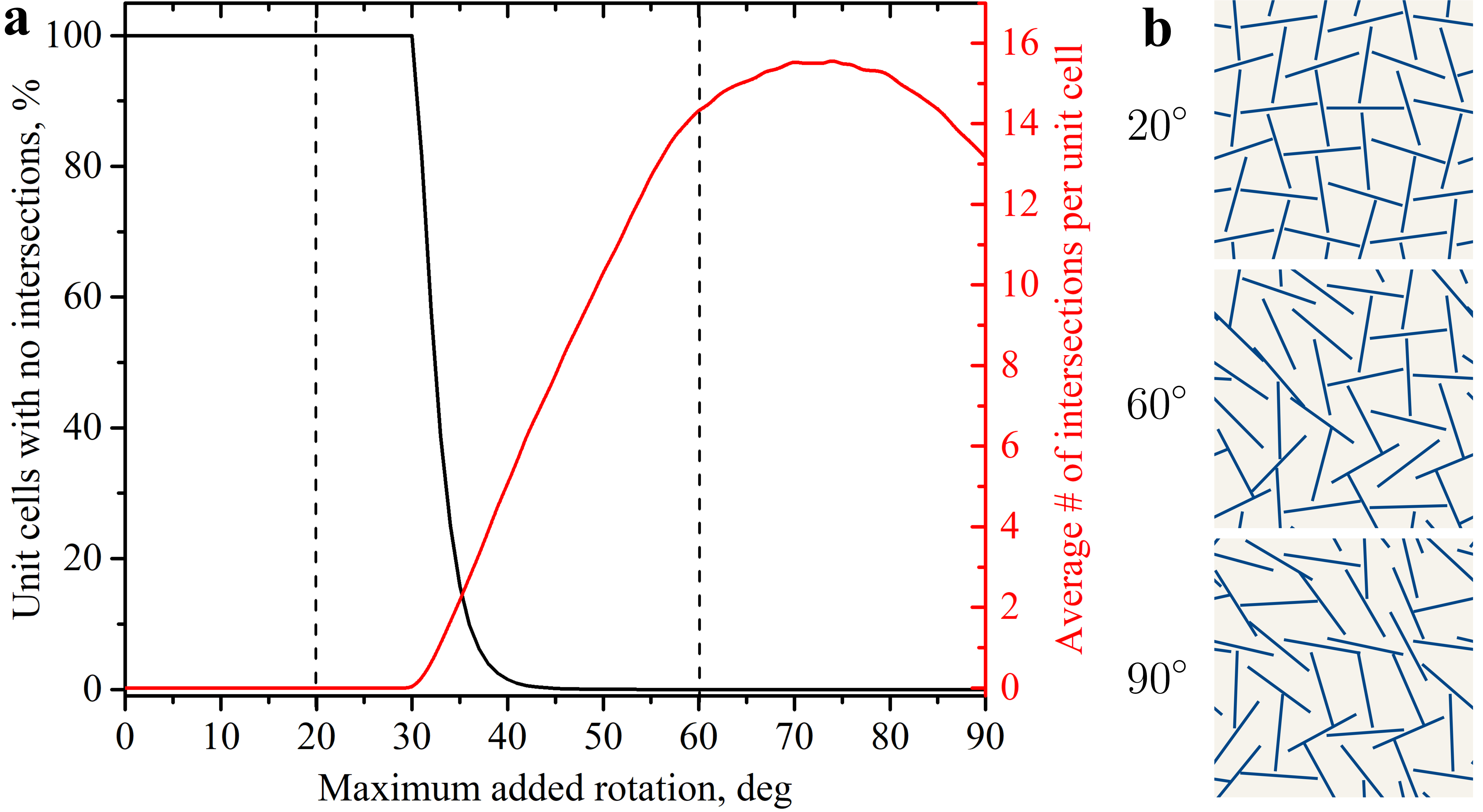}             
		\caption{\text{Limiting the perturbations enables control over the success rate of generation.} \textbf{a} The average number of intersections in the samples and the likelihood of generating unit cells without intersections vs maximum deviation $\beta_{max}$ from the base structure. \textbf{b} Exemplary unit cells for a maximum added rotations $\beta_{max}$ of $20^\circ$, $60^\circ$ and $90^\circ$.}
		\label{fig:intersections_by_angle}
\end{figure}

To demonstrate the practical limitations of the generative algorithms discussed in the previous sections, we assessed their ability to generate kirigami metamaterials akin to those introduced by Grima et al. \cite{Grima2016AuxeticPerforated}. We started with the alternating $6\times6$ pattern (Figure~\ref{fig:rotation}a) and introduced random rotations for each cut, maintaining the periodicity of structure in both directions. These added rotations are denoted as $\beta_{i,j}$ for a given cut $c_{i,j}$, where $ i,j=0,..,5$ (Figure~\ref{fig:rotation}b). Correspondingly, the angle of each cut relative to the vertical direction is denoted as $\alpha_{i,j}$. The dimensions were chosen to ensure that intersections begin to occur only if the maximum absolute value for the added rotations $\beta_{max}$ exceeds $30^\circ$. The centers of adjacent cuts are one unit apart, while the length of each cut is $l = \sqrt{3}$. Through variation of $\beta_{max}$, it is possible to control the average number of intersections in the generated data and the likelihood of randomly generating unit cells without intersections, as illustrated in  Figure~\ref{fig:intersections_by_angle}a. It becomes virtually impossible to obtain an admissible configuration by randomly selecting 36 rotation values even if a maximum disturbance is limited to $\beta_{max}=60^\circ$, with only three out of a million samples containing no intersections.

Figure~\ref{fig:intersections_by_angle}a illustrates that if $\beta_{max}$ is set to less than $30^\circ$, intersections are not possible. In this case, ED serves as a suitable metric, and any linear combination of two samples remains within the admissible design space. However, once $\beta_{max}$ exceeds this threshold,  the likelihood of randomly generating intersection-free samples rapidly becomes practically negligible. In general, this means that admissible designs become sparsely scattered in the design space, making it challenging for ED to measure similarity, as there is no guarantee that a linear combination of two samples is admissible anymore. Therefore the extent to which ED can effectively describe the similarity between unit cells for generative models can be indirectly controlled through $\beta_{max}$. Here, we created datasets for three distinct values of $\beta_{max}$: $20^\circ$, $60^\circ$ and $90^\circ$, to see the corresponding effect on different machine learning algorithms. Each dataset comprised $20,000$ unit cells, represented as $6\times6$ matrices containing the values of $\alpha_{i,j}$. Given the near impossibility of randomly generating intersection-free unit cells for $\beta_{max}=60^\circ$ and $\beta_{max}=90^\circ$, we employed a randomization process. Starting from the base alternating structure, each cut was sequentially replaced with another cut, chosen randomly from those that would not create intersections. This sequence of random replacements was repeated until each cut had been replaced 200 times. Examples of admissible designs for $\beta_{max}$ of $20^\circ$, $60^\circ$ and $90^\circ$ are shown in Figure~\ref{fig:intersections_by_angle}b. 

\section{Results}
In order to demonstrate the varying degree of reliance of four different machine learning algorithms (VAE, GAN, WGAN and DDPM) on  ED, generators for each respective approach were trained on three datasets ($\beta_{max}=$ $20^\circ$, $60^\circ$ and $90^\circ$) separately. Theory suggests that VAE and WGAN will learn to avoid intersections only when ED is applicable for the dataset, i.e., when rotations are limited. Proving that an approach is not learning can be conceptually more challenging than showing that one is. The inability to learn might stem from insufficient model complexity or poor parameter selection. To rule out these factors, VAE, GAN and WGAN models had almost identical architectures with consistent hyperparameters across all experiments. This approach ensures that observed differences in performance can be attributed solely to the training process. Thus, if one of the methods successfully learns to avoid intersections, the model complexity and parameter choices are not to blame.  

\subsection{Implementation}
For implementing the four machine learning approaches - VAE, GAN, WGAN, and DDPM - the PyTorch deep learning framework \cite{Paszke2019PyTorch} was chosen. PyTorch provides the capability to capture the periodicity of the samples by using Convolutional Neural Networks (CNN) with circular padding. CNN architecture is particularly well-suited for the unit cells of the kirigami metamaterial under investigation, as intersections can be effectively represented through hierarchical feature mapping. Intersections primarily depend on the direct neighbors of a cut, which corresponds to a 3x3 convolution, with these neighboring arrangements influencing each other. Both GAN architectures, as well as the decoder of the VAE, were based on the DCGAN framework \cite{Radford2015UnsupervisedRL}. Batch Normalization \cite{Ioffe2015BatchNormalization} was employed in VAE's encoder and sparingly in the generators/decoder to minimize batch internal dependencies \cite{Salimans2016ImprovedTechniques}. The discriminator architecture was nearly identical for both GAN and WGAN, with the addition of a sigmoid activation function for the GAN due to different value range requirements imposed by the objective functions. Given the potential instability of GAN training, different learning rates were utilized for the generator and discriminator ($1\text{e}5, 5\text{e}4$) to improve convergence towards a local Nash equilibrium \cite{Heusel2017GAnsTrained}. Due to the restrictions on the dimensions of the latent space, a different architecture was required for DDPM \cite{ho2020denoising}. A convolutional U-Net architecture \cite{Ronneberger2015UNet} was chosen there, with filter numbers similar to the other generative network. For all models, Adam \cite{Kingma2015Adam} was used as an optimizer, and a batch size was set to $32$. Note that no implicit penalties for intersections were given, enabling models to learn the restrictions just by examining the training dataset. 

\begin{figure}[t]
        \centering
    	\includegraphics[width=\textwidth]{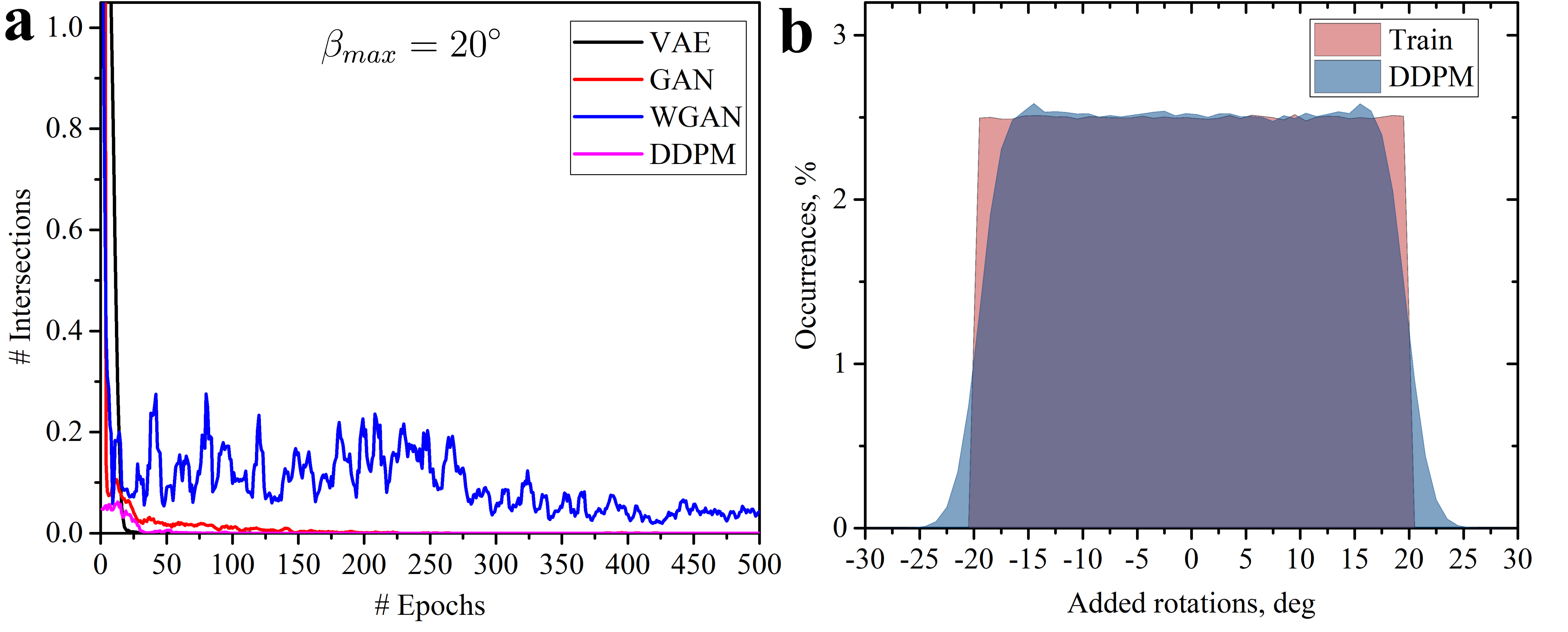}
		\caption{\text{Training of models for $\beta_{max}=20^\circ$}. \textbf{a} The evolution of the average number of intersections during training for unit cells generated by different machine learning approaches. An averaging over five epochs was used for curve smoothing. \textbf{b} Distribution of the cuts with the specific added angles $\beta_{i,j}$ in the training dataset for $\beta_{max}=20^\circ$ (red) and in the set generated by trained DDPM (blue).} 
		\label{fig:20deg}
\end{figure}	

\begin{figure}[ht]
 
        \includegraphics[width=\textwidth]{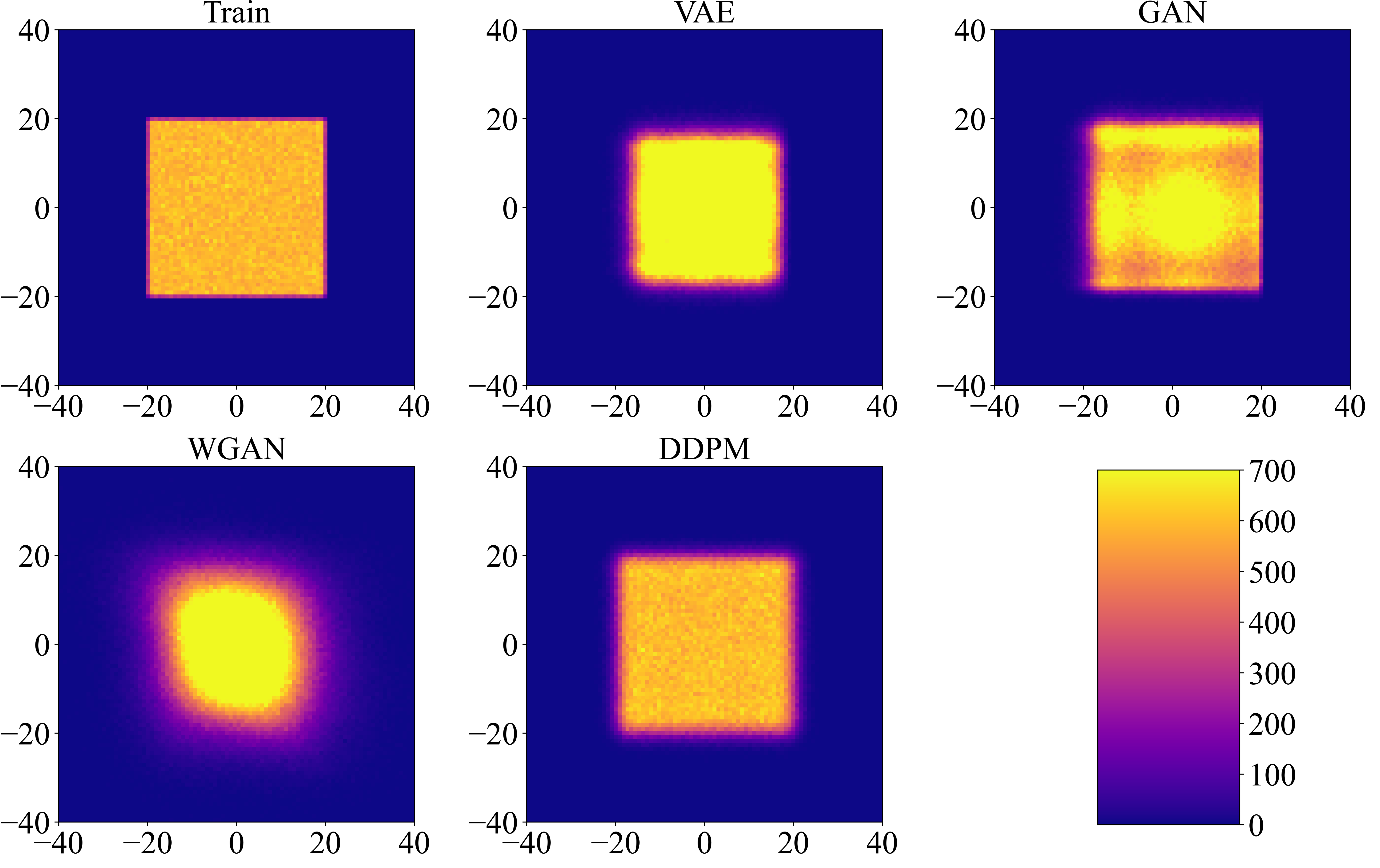}
        \caption{\text{Relation between adjacent cuts for $\beta_{max}=20^\circ$}. 2D histograms show the likelihood of angle combinations for a cut and its bottom neighbor for a maximum absolute value for $\beta=20^\circ$. Only angles at positions that correspond to vertical cuts in the base structure were chosen as the first elements in pairs..}
		\label{fig:dependency20}
\end{figure}	

\subsection{Euclidean Case}
As previously discussed, when the maximum absolute value for the added rotations $\beta_{max}$ is set to $20^\circ$, the angles of these rotations can be chosen independently, allowing ED to effectively measure similarity between samples. This scenario is akin to the classic mechanical metamaterials with benign parameterization. In this case, to generate intersection-free unit cells, the model simply needs to understand that the angle of each cut must be confined within a specific range. Figure~\ref{fig:20deg}a demonstrates that after a few epochs, three out of four approaches learn to generate unit cells with none or very few intersections on average. The WGAN is capable of reducing the number of intersections to less than 0.1 on average, although it exhibits poorer stability during training. In general, if $\beta_{i,j}$ values are drawn from a uniform random distribution with $\beta_{max}=20^\circ$, only admissible configurations are created. The histograms in Figure~\ref{fig:20deg}b reveal that after 3000 epochs, the DDPM successfully learns to emulate this uniform distribution of rotation angles $\beta_{i,j}$, maintaining the generated angles within the range of $-25^\circ$ to $25^\circ$. Minor deviations between the training dataset and the DDPM-generated datasets at the boundary angles of $-20^\circ$ and $20^\circ$ can be attributed to the challenges of learning sharp transitions within continuous models.

Another method to evaluate the effectiveness of a model in learning the intricate constraints of the design space involves examining the overall distribution of rotation angles between neighboring cuts. If the samples generated by the trained model exhibit distributions that closely match those of the training dataset, the model can be deemed suitable for generation. Figure~\ref{fig:dependency20} shows 2D histograms for the training dataset and datasets generated by trained models. The color intensity represents the number of instances where a random cut and its bottom-side neighbor possess a specific combination of angles ($\beta_{i,j}$, $\beta_{i+1,j}$). For illustrative purposes, the first elements of these angle pairs are always from cuts at positions corresponding to vertical cuts in the initial undisturbed sample (Figure~\ref{fig:rotation}a). Since no intersections are possible by construction due to $\beta_{max}= 20^\circ$, all combinations of neighboring angles are equally likely to be observed in the training dataset, as indicated by the homogeneous square in Figure~ \ref{fig:dependency20}. A comparison of the datasets generated by trained models with the initial training dataset reveals that all evaluated models (VAE, GAN, WGAN, DDPM) effectively capture the limits of the perturbations from the initial alternating pattern. However, it is noticeable that the VAE model slightly narrows the range of generated angles, avoiding borderline cases. Simultaneously, the histogram for WGAN clearly shows a more Gaussian distribution rather than a uniform one, indicating the challenge in capturing the angle relationships from the training dataset. The other models (GAN and DDPM) maintain a nearly uniform 2D distribution akin to the training dataset.

\subsection{Fully Random Case}

\begin{figure}[t]
        \centering
    	\includegraphics[width=\textwidth]{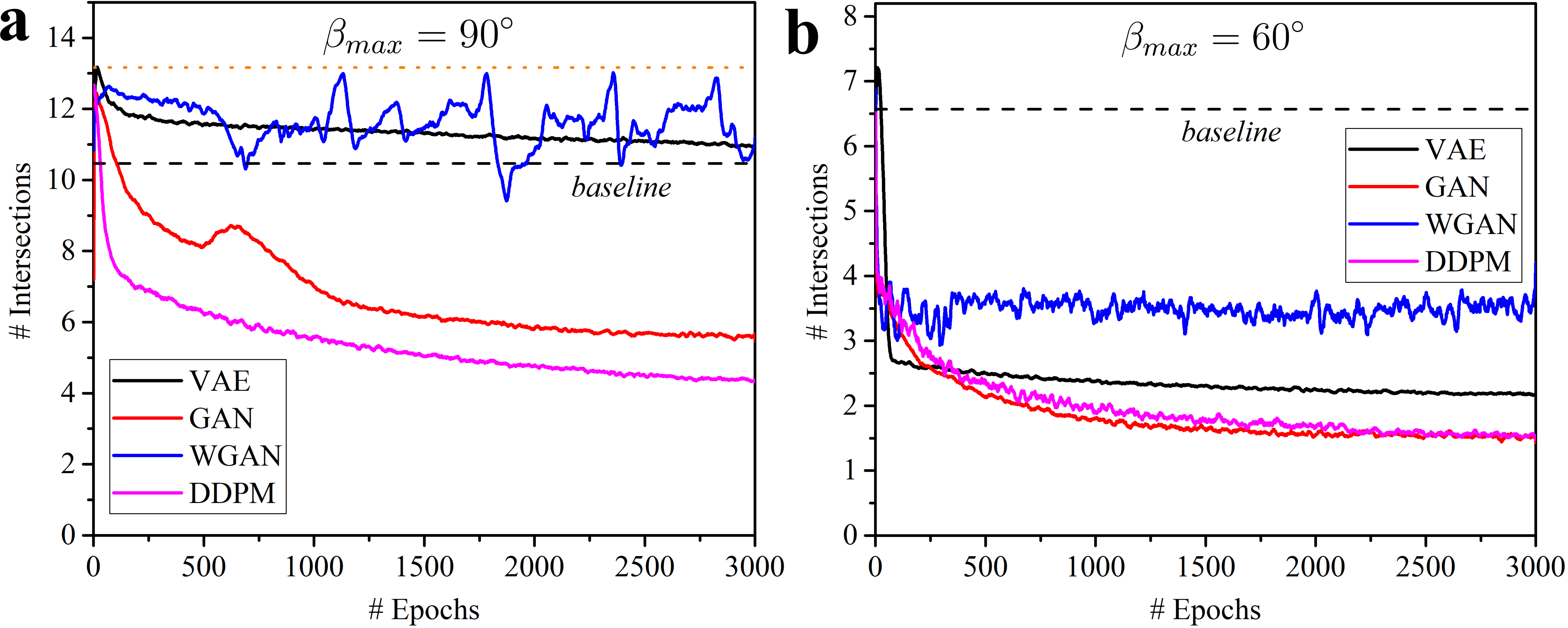}   
		\caption{\text{Training of models for $\beta_{max}=90^\circ$ (a) and $\beta_{max}=60^\circ$ (b)}. The evolution of the average number of intersections during training for unit cells generated by different machine learning approaches. An averaging over 20 epochs was used for curve smoothing. The dashed line corresponds to the average number of intersections assuming random sampling from the training dataset. }
		\label{fig:training90}
\end{figure}	

When the maximal absolute value for added rotations, $\beta_{max}$, is set to $90^\circ$, the design space encompasses all possible samples with non-intersecting cuts. The final rotations of the cuts $\alpha_{i,j}$ are no longer affected by their position in the unit cell and depend only on the rotations of the neighboring cuts instead. This dependency influences the frequency at which certain rotations occur in the dataset, as some rotations are less likely to result in intersections with random neighbors. As a result, randomly generated unit cells, created by sampling cuts based on the angle distribution of the training dataset, typically have fewer intersections (represented by the black dashed line in Figure~\ref{fig:training90}a) compared to those generated from a uniform distribution (indicated by the orange dotted line). Therefore, a reduction in the number of intersections during the training of a machine learning algorithm may occur for two different reasons: the algorithm might learn to fit the angle distribution, or it might also learn the dependency of cuts on each other. In this context, the approach of randomly drawing the rotation angles of the cuts from the angle distribution of the training data serves as a valuable baseline. A decrease in the number of intersections on average during training unequivocally indicates that the model is learning the dependencies between neighboring cuts. 
Figure~\ref{fig:training90}a demonstrates that VAE does not succeed in reducing the average number of intersections below the established baseline. Meanwhile, WGAN manages to slightly lower the number of intersections without going beyond the baseline but fails to converge, which is a common drawback of WGAN \cite{Mescheder2018WhichTraining}. In contrast, both GAN and DDPM achieve significantly lower intersection counts, although they still fall short of generating completely intersection-free samples.

\begin{figure}[t]
        \includegraphics[width=\textwidth]{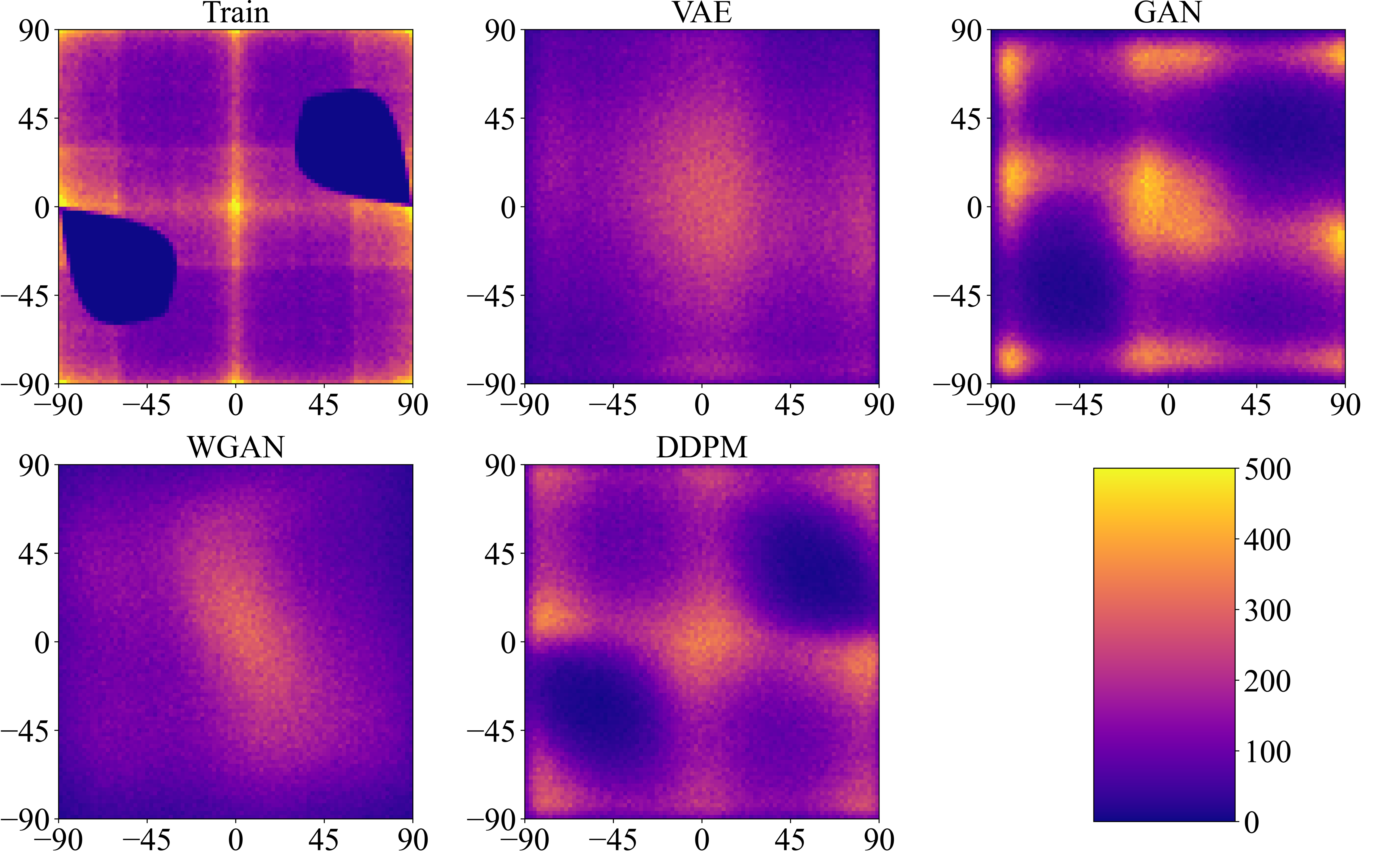}  
        \caption{\text{Relation between adjacent cuts for $\beta_{max}=90^\circ$}. 2D histograms show the likelihood of angle combinations for a cut and its bottom neighbor for fully random rotations of the cuts. Only angles at positions that correspond to vertical cuts in the base structure were chosen as the first elements in pairs.}
		\label{fig:dependency90}
\end{figure}	

A more illustrative measure of whether different generative approaches successfully learn the dependencies between neighboring cuts, as previously mentioned, can be captured using 2D histograms (Figure~\ref{fig:dependency90}). Unlike the dataset with $\beta_{max} = 20^\circ$, the distribution of angles in neighboring cut pairs for the training dataset with $\beta_{max} = 90^\circ$ is no longer uniform. Recall that the dark regions in the histogram for the training dataset correspond to angle pairs where cuts intersect (compare with Figure~\ref{fig:configurations}b). Therefore, avoiding these zones is an important indicator of a successful generative model. By comparing the training dataset with datasets generated by different trained models, we can categorize these models into two groups. Models belonging to the first category (VAE and WGAN) can limit the range of rotation angles but fail to capture the dependency of a cut on its bottom neighbor. Notably, the datasets generated by trained VAE and WGAN models contain angle pairs even from non-admissible zones, likely due to relying on linear interpolation between samples. On the other hand, models from the second category (GAN and DDPM) demonstrate a much better understanding of the design space by learning additional constraints between neighboring cuts, with DDPM slightly outperforming GAN.

\subsection{Intermediate Case}

When the maximum absolute value for the added rotations, $\beta_{max}$, is set to $60^\circ$, it constitutes an intermediate situation between the Euclidean and the fully random cases, with intersections still occurring in the design space. Figure~\ref{fig:intersections_by_angle}a illustrates that it remains almost impossible to generate designs without intersections by chance, even under stricter restrictions compared to the fully random case. Figure~\ref{fig:training90}b displays the training progress for different generative models on the dataset with $\beta_{max}=60^\circ$. Similar to the fully random scenario, all models exhibit gradual improvement during training, as evidenced by a decrease in the average number of intersections in the generated samples.  However, in contrast to the $\beta_{max}=90^\circ$ case, all models are shown to be capable of going significantly below the baseline defined by random sampling from the training dataset distribution. Since the learning of the angle restriction ($|\beta_{i,j}|<60^\circ$) plays an important role, both VAE and WGAN show their capacity for performing that task, similar to the Euclidean case. Nevertheless, as in the fully random case, DDPM and GAN surpass the other two models, more accurately recreating the admissible design space after training. While DDPM does not achieve a 100\% success rate in generating intersection-free configurations, an average of 1.5 intersections per sample is observed. This allows approximately one in every four generated samples to fall within the admissible design space. This represents a more than 50-fold improvement compared to sampling from the training set and a 100,000-fold improvement over uniform distribution sampling. These results underscore the potential of DDPM for generating metamaterials with complex geometrical restrictions. 

While DDPM demonstrates the best performance in this scenario, it is noteworthy that previously weaker models, such as VAE and WGAN, also show significant improvements. Given that their network architectures are identical to those used in the fully random case, it logically follows that changes in the design space are the primary contributors to their improved performance. To avoid the need to work in a 36D space, the corresponding changes in the characteristics of the design space can be illustrated using the example of two neighboring cuts with perturbations added to the initial angles of $0^\circ$ and $90^\circ$ (Figure~\ref{fig:rotation}b). In the case of fully random rotations ($\beta=90^\circ$), the fraction of angle pairs corresponding to intersecting configurations, calculated as the normalized area of non-admissible zones (shown in Figure~\ref{fig:configurations}b), is approximately 16.5\%. This implies that one in six randomly chosen cut pairs is non-admissible. Surprisingly, the probability of intersection under a $\beta_{max}=60^\circ$ constraint increases to 18.7\%. This trend also holds true for $6\times 6$ samples, where the average number of intersections is slightly higher for $\beta_{max}=60^\circ$ as compared to $\beta_{max}=90^\circ$ (Figure~\ref{fig:intersections_by_angle}a). Therefore, the size of the non-admissible zone alone does not account for the improved performance of generative models in the intermediate case. 

At the same time, an alternative metric, closely linked to ED, can be considered. As shown prior, the suitability of ED as a similarity measure is compromised when there is no direct path between samples, as demonstrated in Figure~\ref{fig:configurations}. Therefore, the shapes and positions of non-admissible zones, in addition to their overall area, could significantly influence the appropriateness of ED as a similarity measure in the examined cases. In the previous example involving two adjacent cuts, there is a 23.7\% probability that a straight path connecting two random points within the admissible design space of the fully random case ($\beta_{max}=90^\circ$) passes through a non-admissible zone. However, when $\beta_{max}$ is set to $60^\circ$, this probability drops significantly to only 3.5\%, making ED a much more suitable similarity measure. While generalizing these findings from this 2D example to a 36D case is not straightforward, the observed relationships between neighboring cuts suggest that the improved performance of certain models, particularly VAE and WGAN, in the intermediate case is likely due to a design space that aligns better with ED metric. Consequently, the ability to influence the "goodness" of the design space through the selection of $\beta_{max}$ could hold significant promise for future generative models tailored to deal with the complex design spaces of mechanical metamaterials.

\section{Conclusion}

Machine learning is deeply embedded into the research on mechanical metamaterials, as evidenced by the growing number of successful generative approaches applied to inverse design problems. However, this perceived success may be somewhat misleading, akin to survivorship bias, where only designs with "nice" parameterizations are considered. In image generation, machine learning often derives design constraints from training data, whereas, for many mechanical metamaterials, parameterizations are preselected to include these constraints. Our study extends beyond these benevolent parameterizations to examine more complex kirigami structures. 
One identified issue with the application of generative models to such metamaterials lies in the inapplicability of the classical Euclidean distance (ED) as a metric for assessing the similarity between unit cells. 

Thus, in this study, we explore the extent to which four of the most common generative design algorithms -- VAE, GAN, WGAN, and DDPM -- rely on ED. Using established theoretical findings and the example of kirigami structures, we demonstrate that out of these four algorithms, both VAE and WGAN depend on this similarity measure for effective generation. This dependence limits their ability to learn complex design space constraints, although they are more suitable for generating metamaterials with simpler all-admissible parameterizations due to their stability and lower computational costs during training as compared to GAN and DDPM. In contrast, GAN and DDPM demonstrate potential in learning design space limitations but still fall short of fully capturing these constraints. This suggests that reliance on ED is just one factor contributing to the lack of generative models for kirigami metamaterials, highlighting a need for further investigation. 

One possibility is related to the difficulty of defining sharp decision boundaries from a training dataset due to the extremely low likelihood of encountering edge cases where two cuts barely avoid intersection. Active Learning, which is concerned with choosing samples in a dataset so that the information gain is maximized, might offer the solution. This field has recently seen rapid progress, particularly with the development of Generative flow networks \cite{Bengio2021FlowNetwork}, however these methods still need to be adapted for complex and high-dimensional spaces. 
Another possible factor in the struggle of generative models (GAN and DDPM) with kirigami metamaterials is associated with their mapping of the data to a low-dimensional Euclidean manifold. If such mapping does not exist, and the data lies on a more complex manifold, GAN and DDPM will perform suboptimally. Research into extending diffusion models to address such cases is quite young but promising \cite{DeBortoli2022Advances, Huang2022Riemannian, Zhuang2023diffusion}. 

Our research highlights the inherent limitations of classical generative approaches when employed in the domain of mechanical metamaterials, as opposed to their typical application in image generation. We underscore the reliance of these methods on the Euclidean distance metric, which is unsuitable for many metamaterials with intricate design spaces. The kirigami metamaterials presented here serve as an ideal benchmark for the development of new generative models, given that the complexity of their design space can be modulated by adjusting the perturbations of the initial system.

\section*{Acknowledgments}
Funded by the Deutsche Forschungsgemeinschaft (DFG, German Research Foundation) under Germany’s Excellence Strategy – EXC-2193/1 – 390951807. The authors acknowledge support by the state of Baden-Württemberg through bwHPC and the German Research Foundation (DFG) through grant no INST 39/963-1 FUGG (bwForCluster NEMO).

\section*{Declaration of competing interest}
The authors declare that they have no known competing financial interests or personal relationships that could have appeared to influence the work reported in this paper.

\section*{Data availability}
Data supporting the findings of this study are available from the corresponding author upon reasonable request.

\bibliographystyle{ieeetr}
\bibliography{manuscript}

\end{document}